\documentclass[pre,twocolumn,preprintnumbers,amsmath,amssymb]{revtex4}
\usepackage{amssymb}
\usepackage{latexsym}
\usepackage{epsfig}
\usepackage{color}

\begin{document}

\title{Implications of maximum acceleration on dynamics}

\author{H. Moradpour$^a$\footnote{h.moradpour@riaam.ac.ir}, A. Amiri$^a$, A. Sheykhi$^{b}$\footnote{asheykhi@shirazu.ac.ir}}
\address{$^a$ Research Institute for Astronomy and Astrophysics of Maragha (RIAAM), P.O. Box
55134-441, Maragha, Iran\\
$^b$ Physics Department and Biruni Observatory, College of
Sciences, Shiraz University, Shiraz 71454, Iran}

\begin{abstract}
Considering the corrected Unruh temperature as well as the
entropic force perspective of gravity, we are able to derive the
modification of the Newton's law of gravity. In addition, we also
investigate the effects of the highest achievable acceleration
correction to the Unruh temperature on the Poisson equation.
Moreover, we address modifications to the Newtonian cosmology as
well as the Friedmann equations corresponding to the upper bound
for acceleration.
\end{abstract}


\maketitle
\bigskip
\section{Introduction\label{Intr}}

Since the discovery of black holes thermodynamics in $1970$'s
\cite{B1,H1,H2,B2,D1,u1}, physicists have been speculating that
there should be some deep connection between gravity and
thermodynamics. Indeed, thermodynamics also gets us some
motivations for obtaining various spacetimes
\cite{zhan1,zhan2,rjp}. According to the black hole
thermodynamics, a black hole has an entropy proportional to its
horizon area and a temperature proportional to its surface
gravity, and the entropy and temperature together with the mass of
the black hole satisfy the first law of thermodynamics. The
pioneering work on the direct connection between gravity and
thermodynamics was done by Jacobson \cite{jacob} who disclosed
that the hyperbolic second order partial differential Einstein
equation can be derived by applying the fundamental relation
$\delta Q=T dS$ together with proportionality of entropy to the
horizon area of the black hole. This profound connection between
the first law of thermodynamics and the gravitational field
equations has been extensively observed in various gravity
theories \cite{jacob1,Pad0,Pad1,Pad11,ahepm,ahepm1}. Applying the
thermodynamics laws to dynamics and static horizons, one may
obtain the gravitational field equations and the Friedmann
equations in a wide range of gravity theory
\cite{ahepm1,j1,j2,j3,j4,j5,sheyw1,sheyw2,j8,j9,j10,j11,j12,j13,plbm,j14,prdm,mrea}.
The deep connection between horizon thermodynamics and
gravitational dynamics, help to understand why the field equations
should encode information about horizon thermodynamics. These
results prompt people to take a statistical physics point of view
on gravity.

The great step towards understanding the statistical origin of
gravity, put forwarded by Verlinde \cite{Ver} who claimed that the
laws of gravity are not fundamental and in particular they emerge
as an entropic force caused by the changes in the information
associated with the positions of material bodies. According to
Verlinde, the tendency of a system to increase its entropy leads
to emergence of gravity and spacetime \cite{Ver}. Moreover, if the
distance between test particle and holographic screen is of order
of the Compton wavelength of test mass, then it is assumed that
particle is completely attracted by the system \cite{Ver}. In
addition, applying the first principles, namely, the holographic
principle and the equipartition law of energy, Verlinde derived
Newton's law of gravitation, the Poisson equation, and in the
relativistic regime the Einstein field equations. Similar
argument was also done by Padmanabhan who observed that the
equipartition law of energy for the horizon degrees of freedom
combined with the thermodynamic relation $S = E/2T$ lead to
Newton's law of gravity \cite{Pad2}. Introducing a new proposal
for the entropic force scenario \cite{cai}, Cai et. al., have
obtained the Newtonian cosmology and Friedmann equations using the
entroic force scenario.

It was also argued that entropic force scenario is in
conflict with the observation of ultracold neutrons in the
gravitational field of Earth \cite{prdc}, a result which makes
this hypothesis doubtable \cite{c1}. However, subsequent studies
have shown that the mentioned experiment can not necessarily
reject the entropic origin of gravity \cite{c2,c3,c4}, motivating
physicists to investigate its relation with generalized entropy
formalism \cite{c4}, conservative force notion \cite{c5}, and the
quantum fluctuations of fields \cite{c6}.

It is worth noting that in the entropic force approach towards
gravity, the entropy expression of the holographic screen, the
equipartition law of energy and the Unruh temperature formula on
the holographic screen play crucial role. Any modification of any
of these quantities may lead to modified versions of the
gravitational field equations as well as Friedmann equations. For
example, the entropy-area relation can be modified from the
inclusion of quantum effects, motivated from the loop quantum
gravity \cite{LQG,LQG1,LQG2,LQG3,LQG4,LQG5,LQG6,LQG7}. It was
shown that by employing the modified entropy-area relation, one
can derive corrections to Newton's law of gravitation as well as
modified Friedmann equations by adopting the viewpoint that
gravity can be emerged as an entropic force \cite{Sheykhi1}.
Moreover, inspired by the Debye model for the equipartition law of
energy in statistical thermodynamics, it was shown that by
modification of the equipartition law of energy in the very low
temperature and adopting the viewpoint of the entropic force,
Einstein field equations and Poisson equation can be modified as
well \cite{SK}. Interestingly, it was also shown that the origin
of the modified Newtonian dynamics (MOND) theory can be understood
from Debye entropic gravity perspective \cite{SK,Gao}. It has also
been shown that if one considers either quantum statistics or
non-extensive statistical mechanics instead of the classical
statistics, then one can find theoretical origins for the MOND
theory \cite{EP,MSCG}. It has also been shown that the
entropic force hypothesis can be used to obtain various
gravitational theories and their corresponding cosmology
\cite{Cai41,Smolin,Li,Tian,Myung1,Vancea,Modesto,BLi,Sheykhi2,Sheykhi21,Sheykhi22,Sheykhi23,Ling,Sheykhi24,Gu,Miao1,
other,mann,SMR,EN,m01,m02,m03,m04,m05,m06,m07,MS,mijtp,m08,m09,m10,m13,m11,EN1,m12}.
It is worthwhile mentioning that none of these attempts study the
effects of corrections to the Unruh temperature on systems.

There are various arguments which claim that there is an upper
bound for the acceleration of systems
\cite{cas1,cas2,cas3,cas4,cas5,cas6,cas7}. This upper bound can
have different values depending on the physical situations of the
system \cite{ijmpd}. These various proposed values for the upper
bound are comparable with the value of the Planck acceleration
which is of order $10^{51}{m}/{s^2}$. It has also been shown that
the quantum value of maximum acceleration is in relation with the
rate of the universe expansion during the Planck era \cite{casco}.
Moreover, this upper bound can modify Unruh temperature
\cite{mpla}. It is useful to note here that there are also another
corrections to Unruh temperature found in
Refs.~\cite{rez1,rez,Jiang,m1,m2,m3}. Taking into account the fact
that the Unruh temperature plays a crucial role in the entropic
force scenario \cite{cai}, one can ask which modifications to the
Newtonian cosmology and Friedmann equations are allowed by the
modified Unruh temperature given in ~\cite{mpla}? Moreover,
whether or not this modified Unruh temperature leads to
corrections to the Newton's law of gravity and the Poisson
equation?

In the present paper, we would like to address the above questions
by taking into account the maximum acceleration correction to
Unruh temperature on the holographic screen. The rest of this
paper is organized as follows. In the next section, we show that
correction to the Unruh temperature modifies Newton's law of
gravity and the Poisson equation. Modifications to the Newtonian
cosmology and the Friedmann equations caused by the corrected
Unruh temperature are explored in section III. We summarize our
results in section IV.
\section{Modification to Newton Law's of gravity and Poisson equation}

Consider a particle with acceleration $a$ moving in a spacetime
with line element $ds^2$. It has been shown that if the particle
acceleration value is bounded by an upper value ($a_m$), then the
line element will be modified as \cite{cas1,cas2}

\begin{eqnarray}
d\tilde{s}^2=\left(1-\frac{a^2}{a_m^2}\right)ds^2.
\end{eqnarray}

\noindent Using the above correction to the line element,
Benedetto and Feoli have proven that the Unruh temperature is also
affected by the maximum acceleration ($a_m$) as \cite{mpla}

\begin{eqnarray}\label{T1}
T=\frac{T_U}{\sqrt{1-\frac{a^2}{a_m^2}}},
\end{eqnarray}

\noindent where

\begin{eqnarray}\label{T00}
T_U=\frac{\hbar a}{2\pi c k_B},
\end{eqnarray}

\noindent is the well-known Unruh temperature \cite{u1}. It is
apparent that if we do not consider the maximum acceleration
limitation (or equally $a_m\rightarrow\infty$), then we have
$d\tilde{s}^2\rightarrow ds^2$ and $T\rightarrow T_U$, desired
results. Finally, it is worth to mention that different values of
$a_m$ have been proposed in various physical situations (see
\cite{ijmpd} and references therein).

We consider a system with total relativistic energy $E=Mc^2$ and finite boundary which forms a closed
surface, and plays the role of storage device for
information, i.e. a holographic screen \cite{Ver}. We also assume that the
holographic principle holds, i.e. the number of surface degrees of freedom ($N_S$) is equal to those of the system bulk ($N_b$) leading to $N_S=N_b\equiv N$. Now, the equipartition law of energy can be employed to find \cite{Ver}

\begin{eqnarray}\label{e1}
E=\frac{1}{2}Nk_BT,
\end{eqnarray}

\noindent where

\begin{eqnarray} \label{N}
N=\frac{A}{\ell_p^2}, \  \   \    \     \  \ \ell_p=\sqrt{\frac{G
\hbar}{c^3}},
\end{eqnarray}

\noindent and $\ell_p$ is the Planck length. It is also useful to note here that
the mass $M=E/c^2$ is located in the center of the holographic screen \cite{Ver}.

\subsection{Correction to the Newton's law of gravity}

We consider a spherical holographic screen with radius $r$ as the
boundary of the system. The area of this sphere is $A=4\pi r^2$.
Combining Eqs.~(\ref{T1})-(\ref{N}) and using relation $E=Mc^2$,
one can obtain

\begin{eqnarray}\label{e3}
\frac{a}{\sqrt{1-\frac{a^2}{a_m^2}}}=\frac{GM}{r^2},
\end{eqnarray}

\noindent which can also be rewritten as

\begin{eqnarray}\label{e30}
a=\frac{a_m}{\sqrt{1+(\frac{a_m}{a_N})^2}},
\end{eqnarray}

\noindent where $a_N\equiv\frac{GM}{r^2}$ is the ordinary
Newtonian acceleration. It is easy to obtain that if $a_N\ll a_m$
($a_m\ll a_N$), then we have $a\simeq a_N$ ($a\simeq a_m$) meaning
that $a_N\leq a\leq a_m$. Thus, modification to the acceleration
may play role in the highly accelerated systems for which $a_N$ is
at least comparable with $a_m$. It is also useful
to mention here that by using Eqs.~(\ref{T1}) and~(\ref{e3}), one
can get $T=\frac{\hbar a_N}{2\pi c k_B}$ as the Unruh temperature
of a holographic screen with radius $r$ felt by an observer with
acceleration $a_N$.

In this way we obtained the correction to the Newton's law of
gravity resulting from the corrections to the  Unruh temperature.
Let us compare the result obtained here with the modified
Newtonian dynamics (MOND) resulting from Deby entropic gravity
\cite{SK}. It was argued that by adopting Debye correction to the
equipartition law of energy in the framework of entropic gravity
scenario, it is possible to understand the theoretical origin of
the MOND theory \cite{Gao,SK}. According to the MOND theory, the
Newton's law of gravity is modified as,

\begin{equation}
\label{MOND} a \mu \left(\frac{a}{a_0}\right)=\frac{G M}{r^2},
\end{equation}

\noindent in order to explain the flat rotational curves of spiral
galaxies \cite{Milgrom,Milgrom1,Milgrom2}. Here $\mu=1$ for
usual-values of accelerations and $\mu=\frac{a}{a_{0}}$($\ll 1$)
if the acceleration '$a$' is extremely low, lower than a critical
value $a_{0}=10^{-10}$ $m/s^{2}$ \cite{SK}. At large distance, at
the galaxy out skirt, the kinematical acceleration '$a$' is
extremely small, smaller than $10^{-10}$ $m/s^{2}$ , i.e., $a\ll
a_{0}$, hence the function $\mu(\frac{a}{a_{0}})=\frac{a}{a_{0}}$.
Consequently, the velocity of star on circular orbit from the
galaxy-center is constant and does not depend on the distance; the
rotational-curve is flat, as it is observed.

Therefore, we conclude that the MOND theory is the modification of
Newton's law of gravity for small acceleration (temperature),
$a\ll a_0=10^{-10}$ $m/s^{2}$, while the modification derived in
Eq.~(\ref{e3}), resulting from correction to Unruh temperature,
will only be tangible for highly accelerated systems (for very
high temperature systems in which $a_N$ is comparable with $A$).

\subsection{Corrections to the Poisson equation}

Since for a system of mass $M$, enclosed by surface $A$, we have
$E=Mc^2$, Eq.~(\ref{e1}) can be written as \cite{Ver}

\begin{eqnarray}\label{m0}
M=\frac{ck_B}{2G\hbar}\int T dA.
\end{eqnarray}

\noindent Moreover, for a system with the acceleration $\vec{a}$, it is
natural to define the Newtonian potential as

\begin{eqnarray}\label{np1}
\vec{a}=-\vec{\nabla}\phi.
\end{eqnarray}

\noindent Following Refs.~\cite{Ver,SK}, we can insert
Eq.~(\ref{np1}) into Eq.~(\ref{T00}) to reach at

\begin{eqnarray}\label{T0}
T_U=\frac{\hbar|\vec{\nabla}\phi|}{2\pi c k_B},
\end{eqnarray}

\noindent for the Unruh temperature. Now, using Eqs.~(\ref{T1})
and (\ref{T0}) in order to rewrite Eq.~(\ref{m0}), one can follow
the approach of Refs.~\cite{Ver,SK} to show that

\begin{eqnarray}\label{m01}
M=\frac{1}{4\pi G}\int \vec{\nabla}\cdot\left[\vec{\nabla}\phi
\mathcal{D}(x)\right] dV = \int{\rho (\vec{x}) dV},
\end{eqnarray}

\noindent where $\rho (\vec{x})$ is the energy density and
$\mathcal{D}(x)=\frac{1}{\sqrt{1-x}}$, with
$x={(\vec{\nabla}\phi)^2}/{a_m^2}$. The above equation can be
rewritten as

\begin{eqnarray}\label{neq1}
\vec{\nabla}\cdot[\vec{\nabla}\phi\mathcal{D}(x)]=4\pi G \rho.
\end{eqnarray}

\noindent This is the modified Poisson equation resulting from
corrected Unruh temperature by adopting the the viewpoint that
gravity is an entropic force. For
$\vec{\nabla}\mathcal{D}(x)\simeq0$, we get
$\mathcal{D}(x)\nabla^2\phi=4\pi G \rho$ as the corrected Poisson
equation. Finally, it is useful to note that the standard Poisson
equation ($\nabla^2\phi=4\pi G \rho$) is recovered for $x\ll1$
($\nabla\phi\ll a_m$).

\section{Modified Newtonian and Friedmann cosmology}

We consider a homogenous and isotropic universe described by a
FLRW metric as

\begin{eqnarray}\label{frw}
ds^{2}=-dt^{2}+R^{2}\left( t\right) \left[ \frac{dr^{2}}{1-k r^{2}}%
+r^{2}d\Omega^{2}\right].
\end{eqnarray}

\noindent where $R(t)$ is scale factor, and $k=-1,0,1$ corresponds
to the open, flat and closed universes, respectively \cite{roos}.
The apparent horizon of this spacetime, which is a marginally
trapped hypersurface with vanishing expansion, is given by

\begin{eqnarray}\label{ah}
\mathcal{R}_A=R(t)r_A=\frac{1}{\sqrt{H^2+{k}/{a^2}}},
\end{eqnarray}

\noindent where $r_A$ is the co-moving radius of the apparent
horizon, and can be considered as a proper causal boundary for
this spacetime \cite{Hay2,Hay22,Bak,sheyw1,sheyw2}. Moreover,
$H={\dot{R}}/{R}$, where dot denotes derivative with respect to
time, is the Hubble parameter. Now, consider a situation in which
the FLRW universe is filled by an energy-momentum source of
$T^{\nu}_{\ \mu}=\textmd{diag}(-\rho,p,p,p)$, where $\rho$ and $p$
denote the energy density and pressure of the cosmic fluid, respectively, which
obeys the energy-momentum conservation law as

\begin{equation}\label{cont}
\dot{\rho}+3H(\rho+p)=0.
\end{equation}

\noindent Since the total mass ($M$) and the active gravitational
(Tolman-Komar) mass ($\mathcal{M}$) confined by the volume $V$ are
evaluated as \cite{cai}

\begin{eqnarray}\label{tm}
M=\int (T_{\mu\nu} u^\mu u^\nu) dV,
\end{eqnarray}

\noindent and

\begin{eqnarray}\label{tkm}
\mathcal{M}=2\int (T_{\mu\nu}-\frac{1}{2}Tg_{\mu\nu}) u^\mu u^\nu
dV,
\end{eqnarray}

\noindent respectively, simple calculations lead to

\begin{eqnarray}\label{tm1}
M=\rho V
\end{eqnarray}

\noindent and

\begin{eqnarray}\label{tkm1}
\mathcal{M}=(\rho+3p)V,
\end{eqnarray}

\noindent for the above definitions of mass \cite{cai}. In order
to obtain the above relations, we assumed $\rho$ and $p$ are
functions of time, an assumption which comes from the homogeneity
and isotropy of universe compatible with the FLRW model of
universe \cite{cai}.


\subsection{Modified Newtonian Cosmology}

Consider the apparent horizon as the holographic surface, we have

\begin{eqnarray}\label{acc}
a=-\frac{d^2\mathcal{R}_A}{dt^2}=-\ddot{R} r_A,
\end{eqnarray}

\noindent for the acceleration of a radially co-moving observer at
$r_A$. Since Eq.~(\ref{T1}) is the backbone of our calculations in
this subsection and the following subsection, our results are
valid for systems in which $|a|<a_m$ parallel to the
$|\ddot{R}|<\frac{a_m}{r_A}$ condition.

Now, inserting Eq. (\ref{acc}) into Eq.~(\ref{e3}), and using~ Eq.
(\ref{tm1}), one obtains the modified Newtonian cosmology as

\begin{eqnarray}\label{12}
\frac{\ddot{R}}{R}\left[1-\left(\frac{\ddot{R}r_A}{a_m}\right)^2\right]^{-{1}/{2}}=-\frac{4\pi
G}{3}\rho.
\end{eqnarray}

\noindent It is apparent that the Newtonian cosmology is recovered
in the absence of the maximum acceleration limitation (or equally
at the $a_m\rightarrow\infty$ limit).

For $|\ddot{R}|\ll\frac{a_m}{r_A}$, we can easily expand this
result to find

\begin{eqnarray}\label{13}
\frac{\ddot{R}}{R}\left[1+\frac{1}{2}\left(\frac{\ddot{R}}{R}\frac{\mathcal{R}_A}{a_m}\right)^2\right]=-\frac{4\pi
G}{3}\rho,
\end{eqnarray}

\noindent where we neglected the higher order terms and used the
$\mathcal{R}_A=R(t)r_A$ relation to obtain this equation.
Therefore, the first modification to the Newtonian cosmology, due
to the maximum acceleration limitation, is in the form of
$O\big[(\frac{\ddot{R}}{R})^3\big]$, and it only plays role in
highly accelerated systems.

\subsection{Modified Friedmann equations}

Here, we are going to find the effects of corrected Unruh
temperature~(\ref{T1}) on the Friedmann equations. In order to
achieve this goal, we use Eq. (\ref{tkm1}) instead of~Eq.
(\ref{tm1}), and follow the recipe which led to~Eq. (\ref{12}).
This procedure leads to

\begin{eqnarray}\label{16a}
\frac{\ddot{R}}{R}\left[1-\left(\frac{\ddot{R}r_A}{a_m}\right)^2\right]^{-{1}/{2}}&=&-\frac{4\pi
G(\rho+3p)}{3}.
\end{eqnarray}

\noindent It is easy to check that the standard second Friedmann
equation is obtainable in the appropriate limit of
$A\rightarrow\infty$. In fact, the corrected term will be tangible
whenever $\ddot{R}r_A$ is comparable with $a_m$. In the
$\frac{\ddot{R}r_A}{a_m}\ll1$ limit, this equation is reduced to

\begin{eqnarray}\label{16ar}
\frac{\ddot{R}}{R}+\frac{1}{2}\left(\frac{\ddot{R}}{R}\right)^3\left[\frac{\mathcal{R}_A}{a_m}\right]^2&=&-\frac{4\pi
G(\rho+3p)}{3}.
\end{eqnarray}

\noindent Indeed, the second term in the lhs is the first order
correction to the acceleration equation due to the maximum
acceleration limitation.

Using Eq.~(\ref{cont}), the rhs of this equation can be written as

\begin{eqnarray}\label{rhs}
-\frac{4\pi G(\rho+3p)}{3}=\frac{4\pi G}{3R}\frac{d(\rho
R^2)}{dR}.
\end{eqnarray}

Combining this result with Eq.~(\ref{16a}), we can get the
modified Friedmann equation as

\begin{eqnarray}\label{16a1}
\frac{1}{R^2}\int\frac{d\dot{R}^2}{\sqrt{1-[\frac{\ddot{R}
r_A}{a_m}]^2}}+\frac{\beta}{R^2}&=&\frac{8\pi G}{3}\rho,
\end{eqnarray}

\noindent where $\beta$ is the integration constant. Since the
first Friedmann equation of the standard cosmology should be
recovered at $a_m\rightarrow\infty$ limit, one easily finds
$\beta=k$, which finally leads to

\begin{eqnarray}\label{16a12}
H^2+\frac{1}{2a_m^2R^2}\int\left(\frac{\ddot{R}}{R}\right)^2\mathcal{R}_A^2d\dot{R}^2+\frac{k}{R^2}&=&\frac{8\pi
G}{3}\rho,
\end{eqnarray}

\noindent in which $H\equiv {\dot{R}}/{R}$ is the Hubble
parameter. We considered the $\frac{\ddot{R}r_A}{a_m}\ll1$ limit,
and expanded the integral function up to the first order of
approximation to obtain this result. Therefore, this equation
together with Eq.~(\ref{16a}) cover the effects of maximum
acceleration limitation on the evolution of FLRW universe. It is
worth noting that the effects of the obtained corrections to the
Friedmann equations may be seen in highly accelerated systems.

\section{Conclusion}

Adopting the Cai et al.'s version of the entropic force scenario
\cite{cai} and taking into account the maximum achievable
acceleration correction to the Unruh temperature
\cite{cas1,cas2,cas3,cas4,cas5,cas6,cas7,ijmpd}, we showed that
the Newton's law of gravity is modified as well. In addition,
following the entropic force scenario and considering a continuous
distribution of density $\rho$, we investigated the effects of the
corrected Unruh temperature on the Poisson equation.

The effects of the maximum acceleration correction to the Unruh
temperature on the Newtonian cosmology have also been addressed.
Finally, we obtained the corrections to the Friedmann equations,
and set the obtained constant ($\beta$) by considering the
original Friedmann equations (standard cosmology) limit.

\acknowledgments{We would like to thank the anonymous reviewers
for their valuable comments. The work of H. Moradpour has been
supported financially by Research Institute for Astronomy \&
Astrophysics of Maragha (RIAAM). A. Sheykhi thanks Shiraz
University Research Council.}

\end{document}